\documentclass[aps,prb,preprint]{revtex4}
\usepackage{graphicx} 

\begin{document}

\author{W. J. Mullin}
\author{J. P. Fern\'{a}ndez}
\affiliation{Department of Physics, University of Massachusetts,
Amherst, Massachusetts 01003}
\title{Bose-Einstein Condensation, Fluctuations, and Recurrence
Relations in Statistical Mechanics}

\begin{abstract}
We calculate certain features of Bose-Einstein condensation in the
ideal gas by using recurrence relations for the partition function. 
The grand canonical ensemble gives inaccurate results for certain
properties of the condensate that are accurately provided by the
canonical ensemble.  Calculations in the latter can be made tractable
for finite systems by means of the recurrence relations.  The ideal
one-dimensional harmonic Bose gas provides a particularly simple
and pedagogically useful model for which detailed results are easily
derived.  An analysis of the Bose system via permutation cycles yields
insight into the physical meaning of the recurrence relations.
\end{abstract}
\maketitle

\section{INTRODUCTION}

The achievement of Bose-Einstein condensation (BEC) in alkali
gases\cite{BEC} was a remarkable feat in atomic and low-temperature
physics.  The gases are most often trapped magnetically in potentials
accurately approximated by harmonic oscillator wells.  The result has
been a deluge of theoretical papers on BEC in harmonic potentials,
both for ideal and interacting gases.\cite{Pethick} Experimentally,
the gases are sufficiently dilute and weakly interacting that the
ideal gas is a good first approximation for their description, making
the subject more accessible to students even in their first
statistical-physics course.

A basic problem with the standard presentation of BEC is that the
grand canonical ensemble misrepresents several physical quatities when
a condensate is present.  For example, grand canonical ensemble
calculations greatly overestimate the fluctuations of the condensate
number.  Although, because of their isolation, the most realistic
description of the experimentally condensed gases is via the
microcanonical ensemble, the canonical ensemble gives equally accurate
results.  In the latter the system of interest is in contact with a
heat bath but the particle number is kept fixed, which is crucial.

Calculations using the canonical ensemble are avoided in most
elementary treatments of BEC because of mathematical complications. 
The grand canonical ensemble removes these complications by putting
the system in contact with a particle bath.  Unfortunately, when there
is a condensate, the de Broglie wavelength can be larger than the
system size, making the distinction between system and bath
meaningless and leading to the fluctuation inaccuracy mentioned. 
Moreover, the trapped Bose systems of current interest consist of
relatively few particles.  There exist simple methods using recurrence
relations that have been exploited often in recent work to treat
finite systems in the canonical ensemble; however, these relations are
not well known outside the research literature.  It is these
techniques that we want to discuss here.

Because the experimental trapped systems are finite, they are not
equivalent to systems in the thermodynamic limit.  The trapped systems
have a nonuniform distribution established by the external harmonic
potential.  The true thermodynamic limit in these systems would
require that as the particle number $N$ is increased, we also would
decrease the frequency $\omega$ of the potential in such a way that
the maximum density---proportional to $\omega^{d}N$, where $d$ is the
spatial dimension of the system---remains constant.\cite{WJM2,EuroLet}
Although this limit is not physically realized and finite systems have
no real phase transitions, the experimental transformation of the
system into its lowest state is still rather sudden. Nevertheless, the
existence of a mathematically sharp phase transition is not crucial to
the description of real systems.  What is important is the appearance
of a ``condensation,'' by which we mean the rapid accumulation of a
substantial fraction of the $N$ particles into the ground state
(without big fluctuations about this average) when the temperature
falls below a certain finite value.  We will show that even a finite
one-dimensional ideal Bose gas in a harmonic potential has this
property.

The choice of the canonical ensemble and the use of the recurrence
relations are particularly suitable for the study of finite
systems. Thus a simple model with physical properties
amenable to calculation is available to be exploited for
pedagogical or other purposes.

In Sec.~\ref{sec:grand} we will review the standard grand canonical
ensemble treatment of the one-dimensional (1D) harmonic ideal Bose gas
and identify a temperature below which there is a substantial
accumulation of particles in the ground state.  We will also identify
the physically unrealistic fluctuations in the ground-state occupation
that appear in the grand canonical ensemble description.  In
Sec.~\ref{sec:bosegas} we will show how the 1D Bose gas can be treated
by developing a recurrence relation for the partition function.  More
general recurrence relations for the average number of particles in a
single-particle state and for the partition function are developed in
Sec.~\ref{sec:recur}.  Armed with these tools, we compare canonical
ensemble calculations with their grand canonical ensemble equivalents. 
For a finite system there are only small differences in the mean
values; however, there are large differences in the root-mean-square
fluctuations.  In Sec.~\ref{sec:cycles} we look at the condensation
problem from a quite different point of view, namely that of
permutation cycles \`{a} la Feynman.\cite{FeynSM} Such a view gives a
physical explanation to several mathematical formulations found in the
previous sections and especially to the partition function recurrence
relation.  From this point of view we also see that the grand
canonical ensemble misrepresents the condensate, while the canonical
ensemble treats it accurately.  We find the somewhat surprising result
that the condensate is made up of equally probable permutation cycles
of all lengths up to the condensate number.

\section{GRAND CANONICAL TREATMENT}\label{sec:grand}

We first consider a grand canonical ensemble treatment of BEC in a
one-dimensional (1D) harmonic well. Although our approach is
typical of most statistical physics textbooks, we know of only
one such book that actually covers this particular
example.\cite{Toda} The harmonic potential leads to equally
spaced single-particle energy levels given by 
\begin{equation}
\epsilon_{p}=p\Delta, \label{ener}
\end{equation}
with $p$ a nonnegative integer. The zero-point energy, omitted in
Eq.~(\ref{ener}), can be restored to any physical quantity at the end
of the calculation. The constant $\Delta $ is related to the harmonic
angular frequency $\omega $ by $\Delta =\hbar \omega$.

In the grand canonical ensemble\cite{Huang} the average number of
particles $N$ is given by the relation 
\begin{equation}
N=\sum_{p}\frac{1}{e^{\beta (\epsilon_{p}-\mu)}-1}, \label{N}
\end{equation}
where $\beta =1/k_{B}T$ and $\mu$ is the chemical potential. In
Bose problems the denominator is often expanded in powers of
$e^{-\beta (\epsilon_{p}-\mu)}$ to yield
\begin{equation}
N=\sum_{l=1}^{\infty} e^{l\beta \mu} \sum_{p=0}^{\infty} e^{-\beta
lp\Delta}=\sum_{l=1}^{\infty} e^{l\beta \mu} Z_{1}(\beta l),
\label{N2}
\end{equation}
where
\begin{equation}
Z_{1}(\beta l)=\frac{1}{1-e^{-\beta l\Delta}} \label{Zone}
\end{equation}
is the one-body partition function at the effective inverse
temperature
$\beta l$. In Sec.~V we will see that the sum over $l$
in Eq.~(\ref{N2}) represents a sum over
permutation cycles.

For the very weak potentials used to trap $N$ particles
experimentally, the harmonic oscillator states are very closely spaced
($\hbar \omega \ll k_{B}T$).  Thus to a good approximation we can
replace the sum over $p$ in Eq.~(\ref{N2}) by an integral. 
(Alternatively, we could directly replace the sum in Eq.~(\ref{N}) by
an integral to arrive at the same result.)  Because $\int \! 
dp\,e^{-\beta lp \Delta}=(\beta l\Delta)^{-1}$, we find
\begin{equation}
N\approx N'=\frac{1}{\beta \Delta} \sum_{l=1}^{\infty} e^{l\beta
\mu} \frac{1}{l}=-\frac{k_{B}T}{\Delta}\ln (1-e^{\beta \mu}).
\label{N3}
\end{equation}
The sum in Eq.~(\ref{N3}) is one of the Bose
integrals,\cite{Boseint} which in this case can be evaluated
analytically. Of course, changing the sum to an integral is valid
only if the summand is a smooth function of $p$; we then lose the
contribution of the lowest state when it becomes occupied with
order $N$ particles. Hence $N'$ in Eq.~(\ref{N3}) is just the
contribution of the excited states, and we obtain the total number
$N$ by including the ground-state population: 
\begin{equation}
N=n_{0}+N',
\end{equation}
where 
\begin{equation}
n_{0}=(e^{-\beta \mu}-1)^{-1}. \label{nzero}
\end{equation}

We want to identify a Bose-Einstein ``transition temperature,''
that is, one below which there will be a sizable fraction of the
$N$ particles in the ground state. From Eq.~(\ref{nzero}) this
requirement implies that
$-\beta
\mu =(\gamma N)^{-1}\ll 1$, where $\gamma$ is a number of order
unity. If we invert Eq.~(\ref{N3}), we obtain
\begin{equation} 1-e^{-\beta N' \Delta}=e^{\beta \mu}\approx
1+\beta \mu,
\end{equation}
so that $\beta N' \Delta \approx \ln\gamma N$.  Because $N'$ is of
order $N$ at the transition, we find that the condensate will be large
for temperatures below $T_{0}$ defined by\cite{WJM2,Ket}
\begin{equation}
T_{0}=\frac{N\Delta}{k_{B}\ln N}. \label{T0}
\end{equation}

We can show\cite{WJM2} that the density of the system is proportional
to $N\Delta$, so that, in the thermodynamic limit, we keep the
numerator $N\Delta $ constant while letting $N\rightarrow \infty$,
$\Delta \rightarrow 0$.  Then the characteristic temperature will go
to zero as 1/$\ln N$, which is small only for extremely large $N$.  In
actual 1D experiments,\cite{Schreck,Gorlitz} where $N$ is about
$10^{4}$, the logarithm reduces this characteristic temperature only
by a factor of order ten compared to $N\Delta/k_{B}$, an
experimentally accessible value.  Nevertheless, we say that there is a
``quasi-condensation'' rather than a real one.  For a two-dimensional
(2D) ideal gas we find that an actual phase transition occurs;
however, in accord with the Hohenberg theorem,\cite{WJM2} this
transition disappears if there are particle-particle interactions.  On
the other hand, some authors have claimed that the 1D and 2D finite
interacting systems at sufficiently low temperature have a \emph{true}
condensation, because the coherence length becomes larger than the
finite condensate size.\cite{Pet,Pet2} (Moreover, in 2D we expect a
true phase transition of the Kosterlitz-Thouless variety\cite{Prok} at
a temperature of order $\Delta\sqrt{N}$.)

It is straightforward to solve Eq.~(\ref{N2}) numerically for the
chemical potential $\mu$ and compute the exact condensate number
Eq.~(\ref{nzero}) for some given average $N$ value.  In Fig.~1 we show
the occupation of the first two energy levels for $N$ = 500.  We see
that $T_{0}$ provides a fairly good estimate of the quasi-condensation
temperature.

Many textbooks compute particle-number fluctuations in the
grand canonical ensemble.\cite{Huang} The grand partition
function for any ideal Bose gas with states $
\epsilon_{p}$ each occupied by $n_{p}$ particles is\cite{Huang}
\begin{equation}
\mathcal{Z}=\sum_{N}Z_{N}\,e^{\beta \mu
N}=\prod_{p}\sum_{n_{p}=0}^{\infty}e^{-\beta
(\epsilon_{p}-\mu)n_{p}}=\prod_{p}\frac{1}{1-e^{-\beta
(\epsilon_{p}-\mu)}}. \label{ZGCE}
\end{equation}
where $Z_{N}$ is the canonical partition function of an $N$-particle
system. The average square deviation of the occupation number $n_{p}$
is
\begin{equation}
\overline{\Delta n_{p}^{2}}\equiv
\overline{(n_{p}-\overline{n}_{p})^{2}}
=(k_{B}T)^{2}\frac{\partial^{2}\ln \mathcal{Z}}{\partial
\epsilon_{p}^{2}}=
\overline{n}_{p} +\overline{n}_{p}^{2}. \label{delnGCE}
\end{equation}
With no condensate we have
$\overline{n}_{p} \ll 1$ for all except a
negligible number of excited states, so $\overline{\Delta
N^{2}}=\sum_{p}
\overline{\Delta n_{p}^{2}}\approx \sum_{p}\overline{n}_{p}=N$,
and we have a normal distribution with
\begin{equation}
\frac{\sqrt{\overline{\Delta N^{2}}}}{N}=O(N^{-1/2}).
\end{equation}
However, with a condensate of order $N$, we have 
\begin{equation}
\frac{\sqrt{\overline{\Delta n_{0}^{2}}}}{N}=O(1),
\end{equation}
that is, the fluctuations of the condensate are as large as the
condensate itself---a manifestly unphysical result.  This problem is
not new,\cite{Fuji,Uhlen} but has received a large amount of recent
attention,%
\cite{Gross1,Gross2,Gross3,Polit,Navez,Weiss,Wilk,Gajda,Lemm,Idzi,Illum}
including the invention of a new ``fourth'' ensemble---the
so-called ``Maxwell demon'' ensemble---to take care of it.\cite{Navez}

There are various explanations of what goes wrong with the grand
canonical ensemble. Grossmann and Holthaus\cite{Gross1} state that
``[T]he relative mean square fluctuations of the ground state
population, and thus the relative fluctuations of the total particle
number, approach unity: as a result of particle exchange with the
reservoir, the uncertainty of the number of particles becomes
comparable with $\langle N\rangle $ itself. This fluctuation
catastrophe is related to the divergency of the quantum coherence
length $\lambda_{T}$ for $T\rightarrow 0$. When $\lambda_{T}$ vastly
exceeds the length scale characterizing the system under
consideration, a rigid distinction between `system' and `reservoir' is
no longer practical.'' The difficulty also can be stated in more
mathematical terms. The grand canonical ensemble often is shown to be
equivalent to the canonical ensemble by using the method of steepest
descents,\cite{terHaar} which evaluates the canonical partition
function by an approximation to a complex integral of the grand
canonical partition function. Wilkens and Weiss\cite{Wilk} state that
``The most common procedure is to evaluate the contour integral in a
stationary phase approximation. In leading order one recovers the
grand-canonical formulation. However, below $T_{c}$, fluctuations are
badly represented in this approach. The reason is that, for large $N$,
the saddle point is located within a distance $O(1/N)$ from the branch
point while the Gaussian approximation for the fluctuations assumes a
much larger range of validity $O(1/N^{1/2})$.''

We will not pursue the cause of this difficulty further, but will
avoid it by using the canonical ensemble.

\section{ONE-DIMENSIONAL BOSE GAS BY THE CANONICAL
ENSEMBLE}\label{sec:bosegas}

We next examine the 1D ideal Bose gas by using the canonical
ensemble. There is a simple recurrence relation for the partition
function of this model. As we will see in Sec.~\ref{sec:recur}, there
are more general recurrence relations by which other problems (for
example, the 3D ideal Bose gas) can be treated. The connection between
the two formulations turns out to be a special case of a famous
theorem in number theory as we demonstrate in Sec.~\ref{sec:recur}.

If we let $z=e^{\beta \mu}$, the grand partition function of
Eq.~(\ref{ZGCE}) can be written as
\begin{equation}
\mathcal{Z}=\sum_{N}Z_{N}z^{N}=\sum_{N}\sum_{\{n\}}{}'
z^{N}e^{-\beta
\Delta \sum_{p}pn_{p}}, \label{ZZ}
\end{equation}
where the prime in the sum over $\{n\}$ implies a sum over all
$n_{0},n_{1},n_{2},\ldots =0,1,2,\ldots$ such that
$\sum_{p}n_{p}=N$. If we define $x=e^{-\beta \Delta}$, the last
exponential in Eq.~(\ref{ZZ}) can be written as $x^{M}$ with
$M=E/\Delta =\sum_{p}p\,n_{p}$ and $\mathcal{Z}$ becomes 
\begin{equation}
\mathcal{Z}=\sum_{N}\sum_{\{n\}}{}'
x^{M}z^{N}=\sum_{N}\sum_{M}c_{N}(M)\,x^{M}z^{N},
\label{Zxz}
\end{equation}
where $c_{N}(M)$ is the degeneracy factor for the energy state $M$ of
$N$ particles.

In the 1D oscillator problem, this degeneracy factor $c_{N}(M)$ has a
very interesting mathematical
property.\cite{Gross3,OldTheory,Temp,Nanda,Schon2} It is just the
number of ways that one can partition the integer $M$ into $N\;$or
less integers. For example, $M=4$ can be partitioned in 5 ways:
$1+1+1+1$, $1+1+2$, $1+3$, $2+2$, and $4$, which is equivalent to the
number of ways that four Bose particles can be put in equally spaced
states 0, 1, 2, 3, 4 to have 4 units of energy. Euler, Gauss, Hardy,
Ramanujan, and many other famous mathematicians have contributed
theorems on partitions.\cite{Andrews}

We do not need an explicit expression for $c_{N}(M)$. We can identify
the canonical partition function in Eq.~(\ref{Zxz}) as
\begin{equation}
Z_{N}=\sum_{M}c_{N}(M)\,x^{M}.
\end{equation}
We also have from Eq.~(\ref{ZGCE}) that 
\begin{equation}
\mathcal{Z}(z)=\prod_{p}\frac{1}{1-zx^{p}}=
\frac{1}{1-z}\frac{1}{1-zx}\frac{1}{1
-zx^{2}}\cdots
\end{equation}
We next replace $z$ in $\mathcal{Z}(z)$ by $xz$ so that 
\begin{equation}
\mathcal{Z}(xz)=\frac{1}{1-zx}\frac{1}{1-zx^{2}}\cdots =
(1-z)\mathcal{Z}(z),
\end{equation}
or 
\begin{equation}
\sum_{N}\sum_{M}c_{N}(M)\,x^{M+N}z^{N}=(1-z)
\sum_{N}\sum_{M}c_{N}(M)\,x^{M}z^{N}.
\end{equation}
We equate equal powers of $z$ to obtain 
\begin{equation}
\sum_{M}c_{N}(M)\,x^{M+N}=\sum_{M}c_{N}(M)\,x^{M}-
\sum_{M}c_{N-1}(M)\,x^{M}
\end{equation}
so that
\begin{equation}
x^{M}Z_{N}=Z_{N}-Z_{N-1},
\end{equation}
and
\begin{equation}
Z_{N}=\frac{1}{1-x^{N}}\,Z_{N-1}. \label{Recur1}
\end{equation}
This recurrence relation is trivial to solve explictly. We obtain
\begin{equation}
Z_{N}=\prod_{k=1}^{N}\frac{1}{1-x^{k}}=\prod_{k=1}^{N}Z_{1}
(\beta k).
\label{Z1DBose}
\end{equation}

Many derivations and uses of this result are found in the current
research
literature.\cite{Gross1,Gross2,Weiss,Wilk,Brosens,Brosens2,Chase} 
although the result itself has been around for many
years.\cite{Toda,OldTheory,Temp,Nanda,Toda2} The above derivation is
from Ref.~\onlinecite{Temp}.  Equation~(\ref{Z1DBose}) is applicable
to the 1D harmonic Bose gas.  However, in a recent article,
Sch\"{o}nhammer\cite{Schon} showed that an almost identical relation
holds for the 1D ideal harmonic \emph{Fermi} gas.  His derivation can
easily be adapted to work for bosons.  The Fermi partition function
differs only in having a factor $e^{-\beta E_{0}}$, where
$E_{0}=N(N-1)\Delta /2$ is the Fermi zero-point energy.  This result
means that the internal energies, given by $-\partial \ln
Z_{N}/\partial \beta$, differ only by $E_{0}$ and that the two systems
have identical heat capacities\cite{OldTheory,Schon2} given by
\begin{equation}
C=\frac{dE}{dT}=k_{B}\sum_{k=1}^{N}\frac{(\beta k\Delta)^{2}e^{\beta
k\Delta}}{(e^{\beta k\Delta}-1)^{2}}.
\label{HtCap}
\end{equation}
As shown in Fig.~2, this quantity is linear in $T$ at low
temperatures (that is, for $\Delta \ll k_{B}T\ll T_{0})$ and
approaches $Nk_{B}$ at
$T\gg T_{0}$. The relation between fermions and
bosons for this system was first pointed out in
Ref.~\onlinecite{OldTheory}.

There is a curious aside to this relation between Bose and Fermi
partition functions.\cite{Toda,May,Aldro,Viefers,Lee} The 1D harmonic
gas has a constant density of states, which leads to the equality of
the heat capacities.  Consider instead \emph{free} particles in 2D
where the single-particle states are $\epsilon_{p}=p^{2}/2m$ and the
density of states is a constant because $p\,dp=m\,d\epsilon$.  One can
show by using standard grand canonical ensemble techniques that the 2D
free Fermi and Bose gases have identical heat capacities.  This result
is rather remarkable considering the considerable difference between
the Fermi and Bose derivations of $C$.  Moreover, the 2D Fermi/Bose
heat capacity is fit extremely well by Eq.~(\ref{HtCap}).

\section{MORE RECURRENCE RELATIONS}\label{sec:recur}

It is possible to go beyond the 1D harmonic case and derive canonical
recurrence relations valid for any ideal gas. We first derive
relations for the distribution functions following a method due to
Schmidt,\cite{Schmidt} who showed how the standard Fermi and Bose
distribution functions in the grand canonical ensemble could be
obtained by this means. We have
\begin{equation}
\overline{n}_{p}(N)\,Z_{N}=\sum_{\{n\}}n_{p}e^{-\beta 
\sum_{k}\epsilon_{k}n_{k}}\delta_{N,\Sigma_{i}n_{i}},
\end{equation}
where the Kronecker delta restricts the sum to $N$ particles. Let
$n_{p}=n_{p}'+1$ and $n_{k}'=n_{k}$ for $k\neq p$. Then 
\begin{equation}
\overline{n}_{p}(N) Z_{N}=\sum_{\{n'\}}(n_{p}'+1)e^{-\beta
(\sum_{k}\epsilon_{k}n_{k}'+\epsilon_{p})}
\delta_{N-1,\Sigma_i^{\vphantom{\prime}} n_{i}'}.
\end{equation}
The term corresponding to $n_{p}'=-1$ does not contribute and the
right side involves standard partition function sums
corresponding to
$N-1$ particles. We have
\begin{equation}
\overline{n}_{p}(N)\,Z_{N}=\left[\overline{n}_{p}(N-1)Z_{N-1}+Z_{N-1}
\right] e^{-\beta \epsilon_{p}},
\end{equation}
or 
\begin{equation}
\overline{n}_{p}(N) =e^{-\beta
\epsilon_{p}}\frac{Z_{N-1}}{Z_{N}}\left[1+
\overline{n}_{p}(N-1)\right] , \label{nrecur}
\end{equation}
which is Schmidt's recurrence relation.\cite{Schmidt}
Equation~(\ref{nrecur}) appeared much earlier in the
literature.\cite{Nanda,Sakai,Ansb,Lands} These derivations assume that
$\overline{n}_{p}(N-1) \approx \overline{n}_{p}(N)$ and use the
relation $Z_{N}=e^{\beta F_{N}}$, where $F_{N}$ is the Helmholtz free
energy and $F_{N}-F_{N-1}\approx \partial F_{N}/\partial N=\mu$, to
find
\begin{equation}
\overline{n}_{p}(N)=\frac{1}{e^{\beta
\epsilon_{p}}Z_{N}/Z_{N-1}-1}=
\frac{1}{e^{\beta (\epsilon_{p}-\mu)}-1}.
\end{equation}
We end up with a canonical derivation of the standard grand
canonical ensemble distribution function for bosons. An analogous
derivation is valid for fermions.\cite{Schmidt,Reif}

Unfortunately, the assumption that $\overline{n}_{p}(N-1) \approx
\overline{n}_{p}(N)$ leads back to the same fluctuation inaccuracies
inherent in the grand canonical ensemble. But we need not make this
assumption; Eq.~(\ref{nrecur}) has a direct solution.  By using the
obvious starting values, $\overline{n}_{p}(0)=0$ and $Z_{0}=1$, we can
prove by induction\cite{Lands} that
\begin{equation}
\overline{n}_{p}(N)=\sum_{l=1}^{N}e^{\beta
\epsilon_{p}l}\frac{Z_{N-l}}{ Z_{N}}. \label{npbar}
\end{equation}
To use this relation, we need the partition functions involved.
If we sum the relation over all $p$, we get 
\begin{equation}
N=\sum_{p}\overline{n}_{p}(N)=\sum_{l=1}^{N}\left[\sum_{p}e^{\beta
\epsilon_{p}l}\right] \frac{Z_{N-l}}{Z_{N}}.
\end{equation}
The quantity in square brackets is just the one-body canonical partition
function at the effective inverse temperature $\beta l$, and a
recurrence relation for $Z_{N}$ results:
\begin{equation}
Z_{N}=\frac{1}{N}\sum_{l=1}^{N}Z_{1}(\beta l)Z_{N-l}. \label{Zrecur}
\end{equation}
This relation was apparently first derived by Landsberg,\cite{Lands}
but appears many times in the current research literature.\cite{Weiss,Wilk,Brosens,Chase,Borr,Tempere,Brosens3,Eck,Borr2,Holz,Pratt,Philippe}
It was derived in this journal by Ford,\cite {Ford} although he made
no application of it.  Ford showed that such a result stems from the
relation of Fermi and Bose partition functions to symmetric
polynomials.  Recently Schmidt and Schnack\cite{Schmidt2} extended
this idea.

The use of Eqs.~(\ref{nrecur}) and (\ref{Zrecur}) allows us to
determine the canonical distribution functions for the finite 1D
harmonic Bose system.  (We could as easily find the properties of the
finite 3D harmonic Bose gas.)  We start the recurrence in
Eq.~(\ref{Zrecur}) with $Z_{0}=1$ and find every $Z_{L}, L\leq N$. 
These values are then put into (\ref{nrecur}), starting with
$\overline{n}_{p}(0)=0$ to find each distribution function $n_{p}(L),
L\leq N$, in sequence.  The results are shown in Fig.~3.  There is a
small disagreement between the results for the canonical ensemble and
the grand canonical ensemble for $N=500$, but these become smaller for
larger $N$.  The real difference between the two ensembles arises in
the fluctuations.

We can, in the same way, develop a recurrence relation for the
mean square distribution. We find
\begin{equation}
\overline{n_{p}^{2}}(N) =e^{-\beta \epsilon_{p}}
\frac{Z_{N-1}}{Z_{N}}
\left[1+\overline{n}_{p}(N-1)+\overline{n}_{p}(N-1)^{2}\right].
\end{equation}
With this relation and Eq.~(\ref{nrecur}) we obtain the
root-mean-square fluctuation in the ground-state distribution
function for the canonical ensemble, which can be compared to the
same result for the grand canonical ensemble as given by
Eq.~(\ref{delnGCE}). The results are shown in Fig.~4. As expected,
we see that the grand canonical ensemble result goes to the total
number of particles as $T$ becomes small, while that of the
canonical ensemble goes to zero after peaking near the
quasi-transition temperature $T_{0}$.

We can sharpen the distinction concerning fluctuations by deriving the
probablility $P_{0}(n)$ of finding $n$ particles in the state with
$\epsilon_{0}=0$. For the 1D harmonic gas in the canonical ensemble
this quantity is given by
\begin{eqnarray}
P_{0}(n) &=&\frac{1}{Z_{N}}\sum_{n_{0},n_{1,} \ldots}e^{-\beta
\Delta 
(n_{1}+2n_{2}+3n_{3}+\cdots)}\delta_{n,n_{0}}
\delta_{N,\Sigma_{i=0}n_{i}}
\nonumber \\
&=&\frac{1}{Z_{N}}\sum_{n_{1},n_{2,}\ldots}e^{-\beta \Delta 
(n_{2}+2n_{3}+3n_{4}+\cdots)}e^{-\beta \Delta
(n_{1}+n_{2}+n_{3}+\cdots)}\delta_{N-n,\Sigma_{i=1}n_{i}}.
\end{eqnarray}
The series in the second exponential on the right can be replaced
by $N-n$ and extracted from the sum; the remaining factor becomes
a partition function for $N-n$ particles or 
\begin{equation}
P_{0}(n)=e^{-\beta \Delta (N-n)}\frac{Z_{N-n}}{Z_{N}}. \qquad
\mbox{(Canonical ensemble)}
\end{equation}
The equivalent quantity can be derived for the grand canonical
ensemble. We find 
\begin{equation}
P_{0}(n)=(1-e^{\beta \mu})e^{n\beta \mu}. \qquad
\mbox{(Grand canonical ensemble)}
\end{equation}
The two $P_{0}(n)$ functions are plotted in Fig.~5 for
$T=0.12T_{0}$. We see that the canonical ensemble value is sharply
peaked around the average value while the grand canonical ensemble
function is incorrectly wide and monotonic.

If we include interactions, the fluctuations in the grand canonical
ensemble are tempered to a more physically reasonable
value.\cite{Polit} However, we would still prefer to be able to treat
the ideal gas correctly for its conceptual and pedagogical importance
and because experimentalists can now reduce the effective interactions
to near zero by use of Feshbach resonances.\cite{Pethick}

One might argue that the microcanonical ensemble, in which the system
of interest is isolated, is more nearly equivalent to the actual
experimental conditions of Ref.~\onlinecite{BEC} than the canonical
ensemble. However, it can be shown by use of the recurrence relations
for the microcanonical ensemble\cite{Weiss,Wilk} that the
microcanonical and canonical ensembles give almost identical results.

\section{PERMUTATION CYCLES}\label{sec:cycles}

The sums in Eqs.~(\ref{N2}) and (\ref{Zrecur}) have a physical
interpretation as sums over permutation cycles. This view of
BEC was first developed by Matsubara\cite{Matsu} and
Feynman,\cite{FeynSM} and was recently discussed in this journal
by one of us.\cite{WJMAJP} Here we examine permutation cycles in
the context of recurrence relations. The boson $N$-body wave
function is symmetric and the density matrix can be written in
terms of symmetrical permutations of particles.\cite{FeynSM} The
partition function is the trace of the density matrix and involves
a sum over all permutations,
\begin{equation}
Z_{N}=\frac{1}{N!}\sum_{\rm{P}} \int \! d\mathbf{r}_{1}\ldots
d\mathbf{r}_{N}\langle 
\mathbf{r}_{P1},\ldots,\mathbf{r}_{PN}|e^{-\beta H}
|\mathbf{r}_{1},\ldots,
\mathbf{r}_{N}\rangle, \label{Zloop}
\end{equation}
where the variable $\mathbf{r}_{P_{j}}$ represents the coordinate
of the particle interchanged with particle $j$ in permutation $P$.

Any $N$-particle permutation can be broken up into smaller
permutation cycles.\cite{FeynSM,WJMAJP} For example, for $N=7$ we
might have a 3-particle permutation cycle $1\rightarrow
2\rightarrow 3\rightarrow 1$ plus a 4-particle cycle $4\rightarrow
5\rightarrow 6\rightarrow 7\rightarrow 4$. The corresponding
matrix element in Eq.~(\ref {Zloop}) breaks up into a product of
cycle terms:
\begin{eqnarray}
\langle 3125674|e^{-\beta H}|1234567\rangle &=&\langle 312|e^{-\beta
H_{3}}|123\rangle \langle 5674|e^{-\beta H_{4}}|4567\rangle 
\nonumber \\ &=&\sum_{p}e^{-3\beta \epsilon_{p}}\sum_{m}e^{-4\beta
\epsilon_{m}}=Z_{1}(3\beta)Z_{1}(4\beta),
\end{eqnarray}
where, for example, $H_{3}=h_{1}+h_{2}+h_{3}$ with
each $h_{i}$ being a one-body Hamiltonian. We have reduced the
cycle matrix elements to one-body partition functions at an
effective temperature. The details of this derivation are given in
Ref.~\onlinecite{WJMAJP}. Every term in the sum of
Eq.~(\ref{Zloop}) can be reduced in this way to a product of
permutation cycles represented by products of one-body partition
functions. A single configuration will consist of
$q_{1}$ loops of length 1, $q_{2}$ loops of length 2, etc., and
may be arranged in $C(q_{1},q_{2},\ldots)$ different ways. Thus we
can write
\begin{equation}
Z_{N}=\frac{1}{N!}\mathop{{\sum}'}_{\{q_{1},q_{2},\ldots \}}
C(q_{1},q_{2},\ldots)\prod_{l}Z_{1}(\beta l)^{q_{l}}, \label{ZN}
\end{equation}
where the prime on the sum implies that it is over all
combinations of permutation cycles such that
\begin{equation}
\sum_{l}q_{l}l=N. \label{restrict}
\end{equation}
Feynman\cite{FeynSM} has given an argument (repeated in
Ref.~\onlinecite{WJMAJP})
to show that
\begin{equation}
C(q_{1},q_{2},\ldots)=\frac{N!}{1^{q_{1}}2^{q_{2}}\ldots
q_{1}!\,q_{2}!\ldots} . \label{count}
\end{equation}
Again there is a connection to the theory of numbers. The sum in
Eq.~(\ref {ZN}) is over the number of ways of partitioning the
integer $N$ into smaller integers. An example of
$C(q_{1},q_{2},\ldots)$ is the breaking of particles 1, 2,
$\ldots, 5$ into a 2-cycle and a 3-cycle, that is, partitioning 5
into $2+3$. With five particles there are several ways of doing
this: We can take particles 1 and 2 in the 2-cycle with 3, 4, and
5 in the 3-cycle, or take particles 1 and 3 in the 2-cycle with
the remaining particles in the 3-cycle, etc. In all there are
$C=5!/(2^{1}3^{1}1!1!)=20$ distinct ways of doing this, as the
reader can verify.

For the case of the 1D harmonic Bose gas, we can combine
Eqs.~(\ref{Zone}), (\ref {Z1DBose}), and (\ref{ZN}) to find an
interesting relation:
\begin{equation}
Z_{N}=\prod_{k=1}^{N}Z_{1}(\beta
k)=\mathop{{\sum}'}_{\{q_{1},q_{2},\ldots \}} 
\frac{1}{
1^{q_{1}}2^{q_{2}}\ldots q_{1}!\,q_{2}!\ldots}\prod_{l}Z_{1}(\beta l)^{q_{l}}
\label{Zperm}
\end{equation}
or
\begin{equation}
\label{43}
\frac{1}{(1-x)(1-x^{2})\cdots (1-x^{N})}=\sum_{{\mbox{\scriptsize\rm{partitions 
of}}\,N}}\frac{1}{1^{q_{1}}2^{q_{2}}\ldots
q_{1}!\,q_{2}!\ldots}\frac{1}{
(1-x)^{q_{1}}(1-x^{2})^{q_{2}}\cdots}.
\end{equation}
Equation~(\ref{43}) is known as Cayley's decomposition in the theory of
partitions of numbers.\cite{Andrews} A simple special case is
$1/(1-x)(1-x^{2})=\frac{1}{2}\left[ 1/(1-x)^{2}+1/(1-x^{2})\right]$.

The right-hand side of Eq.~(\ref{Zperm}) is, in fact, the solution of
the recurrence relation (\ref{Zrecur}).  This solution tells us how to
interpret the recurrence relation itself.  The sum in
Eq.~(\ref{Zrecur}) is a sum over permutation cycles: We can generate
the $N$-body partition function for the Bose system by adding a
particle either as a singlet ($Z_{1}(\beta)$) with the other $N-1$
particles grouped independently ($Z_{N-1}$), or as part of a
pair-exchange cycle ($Z_{1}(2\beta)$) with the other $N-2$ particles
in all their possible combinations ($Z_{N-2}$), or as part of a triple
cycle, and so on, with each configuration having equal probablity
$1/N$.  (The work of Lalo\"{e} et al.\cite{Gruter,CentDens} on
interacting gases is closely related to this approach.)

A further useful quantity is the average number
$\overline{p}_{l}=\overline{q}_{l}l$ of particles involved in
permutation cycles of length $l$. This number is found by using
Eq.~(\ref{Zrecur}):
\begin{equation}
N=\sum_{l}\overline{p}_{l}=\sum_{l}Z_{1}(\beta
l)\frac{Z_{N-l}}{Z_{N}},
\end{equation}
which tells us that 
\begin{equation}
\overline{p}_{l}=Z_{1}(\beta l)\frac{Z_{N-l}}{Z_{N}}. \label{pl}
\end{equation}

We can easily plot $\overline{p}_{l}$, but before we do, it is
useful to separate out the contributions to
$\overline{p}_{l}$ from the condensate and the excited states. From
Eq.~(\ref{npbar}) we have 
\begin{equation}
\overline{n}_{0}=\sum_{l}\frac{Z_{N-l}}{Z_{N}}, \label{ntotperm}
\end{equation}
so that $\overline{p}_{l}^{(0)}=Z_{N-l}/Z_{N}$ is the contribution of
the condensate to the average particle number in the permutation cycle
of length $l$.  However, this quantity is essentially unity until $l$
is of order $N$.  On the other hand, for small $l$ in the
one-dimensional oscillator, $Z_{1}(\beta l) \approx 1/(\beta l\Delta)$
in Eq.~(\ref{pl}) (cf.~Eq.~(\ref{N3})), which is the contribution of
the non-condensate to the permutation cycles.  Figure~6 shows
$\overline{p}_{l}$~vs.~$l$ for the canonical ensemble.  Notice the
rapid drop-off of $\overline{p}_{l}$ for small $l$, corresponding to
the non-condensate.  However, for low temperatures (that is, when
there is a large condensate) $\overline{p}_{l} \approx
\overline{p}_{l}^{(0)}\approx 1$ out to a value equal to the
condensate number $\overline{n}_{0}$, where $\overline{p}_{l}$ must
drop off to satisfy Eq.~(\ref{ntotperm}).  We might have guessed
before the calculation that the condensate consists only of very long
permutation cycles of approximately $\overline{n}_{0}$ particles.  Now
we see that this is not true; the condensate particles have equal
probability of being in singles, pair cycles, triple cycles, and so on
out to an $\overline{n}_{0}$-cycle.

The dotted line in Fig.~6 is the grand canonical ensemble estimate of
$\overline{p}_{l}$.  We find this estimate from Eq.~(\ref{N2}), which
we can show\cite{WJMAJP} also to be a sum over permutation cycles.  Thus
\begin{equation}
\overline{p}_{l}=e^{l\beta \mu}Z_{1}(\beta l).
\qquad \mbox{(Grand canonical ensemble)}
\end{equation}
We also have 
\begin{equation}
n_{0}=\frac{1}{e^{-\beta \mu}-1}=\sum_{l=1}^\infty
e^{\beta \mu l}. \qquad \mbox{(Grand canonical ensemble)}
\end{equation}
The summand $e^{\beta \mu l}$ is the condensate contribution to
$\overline{p}_{l}$ in the grand canonical ensemble. The dotted line in
the plot shows that the grand canonical ensemble does not do a very
good job of representing the true nature of condensate permutation
cycles in the Bose gas.

Because the condensate contribution to $\overline{p}_{l}$ must drop
off at $\overline{n}_{0}$, we could estimate $\overline{n}_{0}$ by
finding the value of $l$ for which $\overline{p}_{l} \approx 0.5$. For
the lowest temperature $0.12T_{0}$ in Fig.~6 this estimate gives
$\overline{n}_{0} = 472$, whereas the exact result is 471. This
approach also allows us to estimate $\overline{n}_{0}$ in
path-integral Monte Carlo simulations involving trapped interacting
particles, where no standard estimators of $\overline{n}_{0}$
exist.\cite{Krauth,HeinMul,MHF}

The picture we have then of the condensate is that it does indeed
fluctuate wildly, \emph{not} in overall particle number as would be
suspected from the grand canonical ensemble, but rather in how it
breaks into permutation cycles, with its particles having an equal
probability of being in cycles of all sizes up to the
condensate number itself.

\section{DISCUSSION}

Our goals in this paper have been multiple: (a)~Illustrate the
inadequacies of the grand canonical ensemble in its depiction of
fluctuations in a Bose-condensed system. (b)~Show how the canonical
ensemble can do a much better job in describing a condensed
system. (c)~Find recurrence relations that allow simple treatments of
finite ideal Bose systems. (d)~Study a simple model system, the 1D
harmonically trapped ideal Bose gas, which illustrates all of the
important elements of BEC in a finite system and has its own
particularly simple recurrence relation.  (e)~Delve deeper into the
intricacies of BEC to find the physical meaning of the recurrence
relations by looking at permutation cycles. (f)~Illustrate the close
relation between Bose and Fermi systems for cases where the density of
states is constant.

All our work here involves non-interacting Bose systems.  Although
real gases have non-negligible physical effects due to interactions,
many of the ideas we have developed carry over to the interacting
regime.\cite{Tempere,Brosens3,Holz,Gruter,CentDens,Krauth,HeinMul,MHF}
Although it is probably not possible to develop recurrence relations
for interacting systems,\cite{LemGroup} the idea of the condensate
involving all sizes of permutation cycles, the usefulness of even a 1D
model of a Bose gas, and the close relation between the 1D Bose and
Fermi gases, are ideas that are still expected to hold for interacting
systems.

An important feature of the present paper is that the 1D model is so
simple that instructors can use it in elementary courses in
statistical physics without sacrificing much important physics.  The
computer programs needed to carry out the recurrence relations are
simple and can be coded by the students themselves.  There are thought
to be few models where the canonical ensemble is soluble, but we have
seen here that any ideal system where the single-particle energies are
known is actually tractable.  Usually one does not care so much about
using the canonical ensemble because the grand canonical ensemble
makes the math easier.  However, we have seen that for the case of
BEC the grand canonical ensemble is not always accurate and the
canonical ensemble becomes not only accessible but necessary.

\begin{acknowledgments}
One of us (WJM) is grateful for many useful conversations about BEC with
Franck Lalo\"{e} and Jean-No\"{e}l Fuchs.
\end{acknowledgments}

\pagebreak
\bigskip
\begin{figure}[!h]
\begin{center}
\includegraphics[width=6.35in]{Fig1.EPSF}
\caption{Grand canonical ensemble calculation of the
number of particles in the two lowest states versus $T/T_{0}$ for
the 1D harmonic Bose gas. The results shown for all
the figures are for $N=500$.}

\end{center}
\label{fig1}
\end{figure}

\pagebreak
\bigskip
\begin{figure}[!h]
\begin{center}
\includegraphics[width=6.35in]{Fig2.EPSF}
\caption{Heat capacity per particle (in units of $k_{B}$) for the 1D
ideal harmonic Bose gas.  This quantity is identical to the same
quantity for a 1D ideal harmonic Fermi gas.\cite{Schon}}
\end{center}
\label{fig2}
\end{figure}

\pagebreak
\bigskip
\begin{figure}[!h]
\begin{center}
\includegraphics[width=6.35in]{Fig3.EPSF}
\caption{Comparison of the grand canonical ensemble and canonical
ensemble calculations of the number of particles in the two lowest
states versus $T/T_{0}$ for the 1D harmonic Bose gas.}
\end{center}
\label{fig3}
\end{figure}

\pagebreak
\bigskip
\begin{figure}[!h]
\begin{center}
\includegraphics[width=6.35in]{Fig4.EPSF}
\caption{Root-mean-square fluctuation of the
number of particles in the ground state of the 1D harmonic Bose gas
for the grand canonical and canonical ensembles.  The fluctuations in
the condensate in the grand canonical ensemble become as large as the
occupation itself, which is unphysical.  The canonical ensemble result
is more reasonable.}
\end{center}
\label{fig4}
\end{figure}

\pagebreak
\bigskip
\begin{figure}[!h]
\begin{center}
\includegraphics[width=6.35in]{Fig5.EPSF}
\caption{The probability of finding $n$ particles in the ground state
versus $n$ for the 1D harmonic Bose gas at $T=0.12T_{0}$ for both the
canonical ensemble (solid line) and grand canonical ensemble (dotted
line).  The result for the grand canonical ensemble is unphysical.}
\end{center}
\label{fig5}
\end{figure}

\pagebreak
\bigskip
\begin{figure}[!h]
\begin{center}
\includegraphics[width=6.35in]{Fig6.EPSF}
\caption{The number of particles in permutation cycles of length $l$
versus $l$ for the canonical ensemble (solid lines) at various
temperatures for the 1D harmonic Bose gas.  Also shown by the dotted
line is the same quantity at the lowest temperature for the grand
canonical ensemble.}
\end{center}
\label{fig6}
\end{figure}


\newpage

\begin{thebibliography}{99}
\bibitem{BEC} M. H. Anderson, J. R. Ensher, M. R. Matthews, C. E.
Wieman, and E. A. Cornell, ``Observation of Bose-Einstein
condensation in a dilute atomic vapor,'' Science \textbf{269},
198--201 (1995); K. B. Davis, M.-O. Mewes, M. R. Andrews, N. J. van
Druten, D. S. Durfee, D. M. Kurn, and W. Ketterle, ``Bose-Einstein
condensation in a gas of sodium atoms,'' Phys. Rev. Lett. 
\textbf{75}, 3969--3973 (1995); C. C. Bradley, C. A. Sackett, J. J.
Tollett, and R. G. Hulet, ``Bose-Einstein condensation of lithium:
Observation of limited condensate number,'' Phys. Rev. Lett.
\textbf{78}, 985--989 (1997).

\bibitem{Pethick} For a review, see C. J. Pethick and H. Smith,
\emph{Bose-Einstein Condensation in Dilute Gases} (Cambridge
University Press, Cambridge, 2002).

\bibitem{WJM2} W. J. Mullin, ``Bose-Einstein condensation in a harmonic
potential,'' J. Low Temp. Phys. \textbf{106}, 615--641 (1997).

\bibitem{EuroLet} K. Damle, T. Senthil, S. N. Majumdar, and
S. Sachdev, ``Phase transition of a Bose gas in a harmonic
potential,'' Europhys. Lett.  \textbf{36}, 6--12 (1996).

\bibitem{FeynSM} R. P. Feynman, \emph{Statistical Mechanics} (W. A.
Benjamin, Reading, MA, 1972)

\bibitem{Toda} M. Toda, R. Kubo, and N. Sait\^{o}, \emph{Statistical
Physics I} (Springer Verlag, Berlin 1983).

\bibitem{Huang} For example, Ref.~\onlinecite{FeynSM} or K. Huang,
\emph{Statistical Mechanics} (John Wiley and Sons, New York, 1987),
2nd ed.

\bibitem{Boseint} F. London, \emph{Superfluids} (Dover, New York, 1964)
Vol. II, p. 203.

\bibitem{Ket} W. Ketterle and N. J. van Druten, ``Bose-Einstein
condensation of a finite number of particles trapped in one or three
dimensions,'' Phys.  Rev.  A \textbf{54}, 656--660 (1996); N. J. van
Druten and W. Ketterle, ``Two-Step Condensation of the Ideal Bose Gas
in Highly Anisotropic Traps,'' Phys. Rev. Lett.  \textbf{79}, 549--552
(1997).

\bibitem{Schreck} F. Schreck, L. Khaykovich, K. L. Corwin, G. Ferrari,
T. Bourdel, J. Cubizolles, and C. Salomon, ``Quasipure Bose-Einstein
Condensate Immersed in a Fermi Sea,'' Phys. Rev. Lett.  \textbf{87},
080403-1--4 (2001).

\bibitem{Gorlitz} A. G\"{o}rlitz, J. M. Vogels, A. E. Leanhardt, C. Raman,
T. L. Gustavson, J. R. Abo-Shaeer, A. P. Chikkatur, S. Gupta, S.
Inouye, T. Rosenband, and W. Ketterle, ``Realization of Bose-Einstein
Condensates in Lower Dimensions,'' Phys.  Rev.  Lett.  \textbf{87},
130402-1--4 (2001).


\bibitem{Pet} D. S. Petrov, M. Holzmann, and G. V. Shlyapnikov,
``Bose-Einstein condensation in quasi-2D trapped gases,'' Phys.  Rev. 
Lett.  \textbf{84}, 2551--2555 (2000).

\bibitem{Pet2} D. S. Petrov, G. V. Shlyapnikov, and J. T. M. Walraven,
``Regimes of Quantum Degeneracy in Trapped 1D Gases,'' Phys.  Rev. 
Lett.  \textbf{85}, 3745--3749 (2000).

\bibitem{Prok} N. V. Prokof'ev, O. A. Ruebenacker, and
B. V. Svistunov, ``Critical point of a weakly interacting
two-dimensional Bose gas,'' Phys. Rev. Lett. \textbf{87}, 270402-1--4
(2001).

\bibitem{Fuji} I. Fujiwara, D. ter Haar, and H. Wergeland, ``Fluctuations
in the population of the ground state of Bose systems,'' J. Stat. Phys. 
\textbf{2}, 329--347 (1970).

\bibitem{Uhlen} R. M. Ziff, G. E. Uhlenbeck and M. Kac, ``The ideal
gas revisited,'' Phys.\ Rep. \textbf{32}, 169--248 (1977).

\bibitem{Gross1} S. Grossmann and M. Holthaus, ``Microcanonical
fluctuations of Bose system's ground state occupation number,''
Phys. Rev. E \textbf{54}, 3495--3498 (1996).

\bibitem{Gross2} S. Grossmann and M. Holthaus, ``Maxwell's demon at
work: two types of Bose condensate fluctuations in power-law traps,''
Opt. Express \textbf{1}, 262--271 (1997); ``Fluctuations of the
particle number in a trapped Bose-Einstein condensate,''
Phys. Rev. Lett. \textbf{79}, 3557--3560 (1997).

\bibitem{Gross3} S. Grossmann and M. Holthaus, ``From number theory to
statistical mechanics: Bose-Einstein condensation in isolated traps,'' Proc.
Heraeus-Sem. 1997, Chaos, Solitons, and Fractals \textbf{10}, 795--804
(1999) or cond-mat/9709045.

\bibitem{Polit} H. D. Politzer, ``Condensate fluctuations of a trapped,
ideal Bose gas,'' Phys. Rev. A \textbf{54}, 5048--5054 (1996).

\bibitem{Navez} P. Navez, D. Bitouk, M. Gajda, Z. Idiaszek, and K. 
Rz\c{a}\.{z}ewski, ``Fourth Statistical Ensemble for the Bose-Einstein 
Condensate,'' Phys Rev. Lett. \textbf{79}, 1789--1792 (1997).

\bibitem{Weiss} C. Weiss and M. Wilkens, ``Particle number counting
statistics in ideal Bose gases,'' Opt. Express \textbf{1}, 272--283 (1997).

\bibitem{Wilk} M. Wilkens and C. Weiss, ``Particle number fluctuations in
an ideal Bose gas,'' J. Mod. Opt. \textbf{44}, 1801--1814 (1997).

\bibitem{Gajda} M. Gajda and K. Rz\c{a}\.{z}ewski, ``Fluctuations of the
Bose-Einstein condensate,'' Phys. Rev. Lett. \textbf{78}, 2686--2689 (1997).

\bibitem{Lemm} L. Lemmens, F. Brosens, and J. Devreese, ``Statistical
mechanics and path integrals for a finite number of bosons,'' Sol. State
Comm. \textbf{109}, 615--620 (1999).

\bibitem{Idzi} Z. Idiaszek, M. Gajda, P. Navez, M. Wilkens, and K.
Rz\c{a}\.{z}ewski, ``Fluctuations of the weakly interacting
Bose-Einstein condensate,'' Phys Rev. Lett. \textbf{82}, 4376--4379
(1999).

\bibitem{Illum} F. Illuminati, P. Navez, M. Wilkens, ``Thermodynamic
identities and particle number fluctuations in weakly interacting
Bose-Einstein condensates,'' J. Phys. B \textbf{32}, L461--L464 (1999).

\bibitem{terHaar} D. ter Haar, \emph{Elements of Statistical Mechanics}
(Rinehart and Co., New York, 1954).

\bibitem{OldTheory} F. C. Auluck and D. S. Kothari, ``Statistical
mechanics and the partitions of numbers,'' Proc.  Camb.  Phil.  Soc. 
\textbf{42}, 272--277 (1946).

\bibitem{Temp} H. N. V. Temperley, ``Statistical mechanics and the
partition of numbers I. The transition of liquid helium,''
Proc. Roy. Soc. London A \textbf{199}, 361--375 (1949).

\bibitem{Nanda} V. S. Nanda, ``Bose-Einstein condensation and the
partition theory of numbers'' Proc. Nat. Inst. Sci. (India)
\textbf{19}, 681--690 (1953).

\bibitem{Schon2} K. Sch\"{o}nhammer and V. Meden, ``Fermion-boson 
transmutation and comparison of statistical ensembles in one dimension,'' 
Am. J. Phys. \textbf{64}, 1168--1176 (1996).

\bibitem{Andrews} G. E. Andrews, \emph{The Theory of Partitions}
(Addison-Wesley, Reading, MA, 1976).

\bibitem{Brosens} F. Brosens, J. T. Devreese, and L. F. Lemmens,
``Canonical Bose-Einstein condensation in a parabolic well,''
Sol. State Comm. \textbf{100}, 123--127 (1996).

\bibitem{Brosens2} F. Brosens, J. T. Devreese, and L. F. Lemmens,
``Thermodynamics of coupled identical oscillators within the path-integral
formalism,'' Phys. Rev. E \textbf{55}, 227--236 (1997).

\bibitem{Chase} K. C. Chase, A. Z. Mekjian, and L. Zamick, ``Canonical
and microcanonical ensemble approaches to Bose-Einstein condensation:
The thermodynamics of particles in harmonic traps,'' Eur. Phys. J. B
\textbf{8}, 281--285 (1999).

\bibitem{Toda2} M. Toda, ``On the relation between Fermions and
Bosons,'' J. Phys. Soc. Japan \textbf{7}, 230 (1952).

\bibitem{Schon} K. Sch\"{o}nhammer, ``Thermodynamics and occupation
numbers of a Fermi gas in the canonical ensemble,'' Am.  J. Phys. 
\textbf{68}, 1032--1037 (2000).

\bibitem{May} R. M. May, ``Quantum Statistics of Ideal Gases in Two 
Dimensions,'' Phys. Rev. \textbf{135}, A1515--A1518 (1964).

\bibitem{Aldro} R. Aldrovandi, ``Two-Dimensional Quantum Gases,''
Fortschr.  Phys.  \textbf{40}, 631--649 (1992).

\bibitem{Viefers} S. Viefers, F. Ravndal, and T. Haugset, ``Ideal
quantum gases in two dimensions,'' Am.  J. Phys.  \textbf{63},
369--376 (1995).

\bibitem{Lee} M. Lee, ``Equivalence of ideal gases in two dimensions
and Landen's relations,'' Phys.  Rev.  E \textbf{55}, 1518--1520
(1997).

\bibitem{Schmidt} H. Schmidt, ``A simple derivation of distribution
functions for Bose and Fermi statistics,'' Am. J. Phys. \textbf{57},
1150--1151 (1989).

\bibitem{Sakai} T. Sakai, ``Gibbs' canonical ensemble and the
distribution law in statistical mechanics,''
Proc. Phys-Math. Soc. Japan \textbf{22}, 193--207 (1940).

\bibitem{Ansb} F. Ansbacher and W. Ehrenberg, ``The derivation of
statistical expressions from Gibbs' canonical ensemble,''
Phil. Mag. \textbf{40}, 626--631 (1949).

\bibitem{Lands} P. T. Landsberg, \emph{Thermodynamics} (Interscience
Publishers, New York, 1961).

\bibitem{Reif} F. Reif, \emph{Fundamentals of Statistical and Thermal
Physics} (McGraw-Hill, New York, 1965). Section 9.3 presents a
somewhat similar derivation of the distribution functions via the
canonical ensemble.

\bibitem{Borr} P. Borrmann and G. Franke, ``Recursion formulas for
quantum statistical partition functions,'' J. Chem. Phys. \textbf{98},
2484--2485 (1993).

\bibitem{Tempere} J. Tempere and J. T. Devreese, ``Canonical Bose-Einstein
condensation of interacting bosons in two dimensions,'' Sol. State Comm. 
\textbf{101}, 657--659 (1997).

\bibitem{Brosens3} F. Brosens, J. T. Devreese, and L. F. Lemmens,
``Correlations in a confined gas of harmonically interacting
spin-polarized fermions,'' Phys. Rev. E \textbf{58}, 1634--1643
(1998).

\bibitem{Eck} B. Eckhardt, ``Eigenvalue statistics in quantum ideal
gases,'' in \emph{Emerging Applications in Number Theory}, edited by
D. Hejkal et al.  (Springer, New York, 1997) pp.\ 187--199.

\bibitem{Borr2} P. Borrmann, J. Hartig, O. M\"{u}lken, and E. R. Hilf,
``Calculation of thermodynamic properties of finite Bose-Einstein
systems,'' Phys. Rev. A \textbf{60}, 1519--1521 (1999).

\bibitem{Holz} M. Holzmann and W. Krauth, ``Transition temperature of the
homogeneous weakly interacting Bose gas,'' Phys. Rev. Lett. \textbf{83},
2687--2690 (1999).

\bibitem{Pratt} S. Pratt, ``Canonical and microcanonical calculations
for Fermi systems, Phys.  Rev.  Lett.  \textbf{84}, 4255--4259 (2000).

\bibitem{Philippe} F. Philippe, J. Arnaud, and L. Chusseau,
``Statistics of non-interacting bosons and fermions in microcanonical,
canonical, and grand-canonical ensembles: A survey,'' math-ph/0211029.

\bibitem{Ford} D. I. Ford, ``A note on the partition function for
systems of independent particles,'' Am. J. Phys. \textbf{39},
215--230 (1971).

\bibitem{Schmidt2} H.-J. Schmidt and J. Schnack, ``Partition functions
and symmetric polynomials,'' Am. J. Phys. \textbf{70}, 53--57 (2002);
``Symmetric polynomials in physics,'' cond-mat/0209397.

\bibitem{Matsu} T. Matsubara, ``Quantum-Statistical Theory of
Liquid Helium,'' Prog. Theor. Phys. \textbf{6}, 714--730 (1951).

\bibitem{WJMAJP} W. J. Mullin, ``The loop-gas approach to
Bose-Einstein condensation for trapped particles,'' Am. J. Phys.
\textbf{68}, 120--128 (2000); ``Permutation cycles in the
Bose-Einstein condensation of a trapped gas,'' Physica B
\textbf{284--288}, 7--8 (2000).

\bibitem{Gruter} P. Gr\"{u}ter, D. M. Ceperley, and F. Lalo\"{e},
``Critical temperature of Bose-Einstein condensation of
hard-sphere gases,'' Phys. Rev. Lett. \textbf{79}, 3549--3552 (1997).

\bibitem{CentDens} M. Holzmann, P. Gr\"{u}ter, and F. Lalo\"{e},
``Bose-Einstein condensation in interacting gases,'' Eur. Phys.
J. B \textbf{10}, 739--760 (1999).

\bibitem{Krauth} W. Krauth, ``Quantum Monte Carlo calculations for
a large number of bosons in a harmonic trap,'' Phys. Rev. Lett.
\textbf{77}, 3695--3698 (1996).

\bibitem{HeinMul} S. Heinrichs and W. J. Mullin, ``Quantum Monte Carlo
calculations for bosons in a two-dimensional harmonic trap,'' J. Low
Temp.  Phys.  \textbf{113}, 231--236 (1998); erratum \textbf{114}, 571
(1999).

\bibitem{MHF} W. J. Mullin, S. Heinrichs, and J. P. Fern\'{a}ndez,
``The condensate number in PIMC treatments of trapped bosons,''
Physica B \textbf{284--288}, 9--10 (2000).

\bibitem{LemGroup} An exception to the statement that recurrence 
relations hold only for ideal systems is the harmonically interacting 
system studied in Refs.~\onlinecite{Brosens2}, \onlinecite{Tempere}, 
and \onlinecite{Brosens3}.

\end{thebibliography}
\end{document}